\pdfoutput=1
\documentclass[11pt,a4paper]{article}

\usepackage[T1]{fontenc}
\usepackage[utf8]{inputenc}
\usepackage[margin=2.4cm]{geometry}
\usepackage{amsmath,amssymb}
\usepackage{graphicx}
\usepackage{booktabs}
\usepackage{siunitx}
\usepackage[hidelinks]{hyperref}
\usepackage{caption}

\DeclareSIUnit{\px}{px}
\DeclareSIUnit{\arcsec}{\arcsecond}

\title{Multi Kernel Gaussian Process Guiding:\\
       Predictive Correction for Mounts with Several Gear Stages}
\author{Rumen G. Bogdanovski\\
        \small INDIGO Astronomy initiative\\
        \small\texttt{rumenastro@gmail.com}}
\date{\today}

\begin{document}
\maketitle

\begin{abstract}
Gaussian process (GP) based predictive periodic error correction, introduced by
Klenske et al.~\cite{klenske2016}, learns a telescope mount's periodic tracking
error online and applies the correction before the error is observed, removing
the reaction lag inherent in purely reactive autoguiders. The published model
uses a single periodic kernel and therefore represents the error of one gear
stage: any repeating shape at one period, harmonics included, but only one
period. Mounts with more than one reduction stage -- strain-wave drives with a
separate input stage, belt-and-worm combinations, transfer gears -- produce
errors at periods that are mutually incommensurate, which beat against each
other and never repeat on a common cycle. We extend the model to a sum over an
arbitrary number of quasi-periodic kernels, one per gear stage, each discovered
from the residual spectrum, gated on the strength of its own spectral line, and
refused if it is commensurate with a period already modelled. We show that a
strictly periodic second kernel is unusable in practice because it demands the
second period to within \SI{0.2}{\percent}, and that a squared-exponential
envelope of about twenty periods removes that sensitivity at a cost of
\SI{0.02}{\px}. On synthetic two-stage errors at realistic periods and
amplitudes the extension reduces the residual by
\SIrange{1}{40}{\percent} against the single-kernel model, while leaving
single-stage cases numerically unchanged. On a recorded, de-noised session
with no second, incommensurate gear frequency, it still
improves on the single-kernel model by \SI{6}{\percent} and on a reactive
hysteresis controller by \SI{45}{\percent}, by absorbing amplitude modulation
of the primary that the single kernel cannot. The implementation is
dependency-free, multiplatform code in the C programming language and is
released as part of the INDIGO project~\cite{indigo}.
\end{abstract}

\section{Introduction}

An autoguider corrects telescope tracking errors by measuring the displacement
of a guide star and issuing compensating mount commands. A purely reactive
controller---proportional, hysteresis, or PI---can only respond to an error
after it has been measured, so its correction always lags the disturbance by at
least one guide cycle. For the dominant component of mount tracking error, the
periodic error of the gear train, that lag is unnecessary: the error is
deterministic and repeats, so it can be predicted.

Klenske et al.~\cite{klenske2016} formalised this as online GP regression on the
accumulated gear error, with a kernel that is the sum of a long squared
exponential (SE) for slow aperiodic wander, an exponentiated-sine-squared (periodic)
kernel for the gear error, and a short SE for transient
wander. The periodic kernel's period is estimated online from the spectrum of
the de-trended residual. Their implementation is available in
PHD2~\cite{phd2}, and a dependency-free reimplementation in the C programming
language forms the \emph{Predictive PEC} mode of the INDIGO guider agent.

A single periodic kernel with period $P$ imposes $f(t) = f(t+P)$ on the
predictable part of the model. This is less restrictive than it first appears:
because the kernel's length scale controls only how fine a feature is resolved
\emph{within} one period, the model can represent an arbitrary repeating
waveform at $P$, including a sharply non-sinusoidal one with many harmonics.
What it cannot represent is two periods whose ratio is irrational. Two gear
stages at, say, \SI{432}{\second} and \SI{155}{\second} produce a sum that
never repeats; forcing it into one periodic kernel leaves the second stage almost
entirely uncorrected.

This paper describes the extension of the model to several gear stages, the
practical problems that arise when a second periodic kernel is added naively,
and the measures that make it usable.

\section{Background: the single-kernel model}
\label{sec:background}

We summarise the model of~\cite{klenske2016} in the notation used below. Let
$y_i$ be the accumulated gear error at gear time $t_i$, reconstructed from the
measured star displacement and the sum of the corrections already applied. The
model is a GP~\cite{rasmussen},
\begin{equation}
  y(t) \sim \mathcal{GP}\!\left(h(t)^{\!\top}\beta,\; k(t,t')\right),
\end{equation}
with an explicit linear trend basis $h(t) = [1, t]^{\!\top}$ absorbing drift,
a per-frame noise variance derived from the guide star's signal-to-noise ratio,
so that a frame measured through cloud is trusted less than a clean one, and
covariance
\begin{equation}
  k(d) \;=\;
  \underbrace{\sigma_0^2 e^{-d^2/2\ell_0^2}}_{\text{long SE}} \;+\;
  \underbrace{\sigma_P^2 \exp\!\left(-\frac{2\sin^2(\pi d/P)}{\ell_P^2}\right)}_{\text{periodic}} \;+\;
  \underbrace{\sigma_1^2 e^{-d^2/2\ell_1^2}}_{\text{short SE}},
  \label{eq:base}
\end{equation}
where $d = t-t'$; Table~\ref{tab:notation} collects the symbols used here and
below. Prediction uses a \emph{projection} kernel that omits the
short SE term: that component exists to absorb short-lived wander during
fitting so it is not misattributed to the gear error, and deliberately does not
propagate into the forecast.

The period $P$ is tracked online. The de-trended, regularised residual is
windowed, its power spectrum computed by FFT, and the dominant line taken as
the period estimate, optionally confined to a band around a user-supplied seed.
The tracked period relaxes towards the estimate with a small learning rate.

Two scalars govern how much the prediction is trusted: a data ramp, which is
the fraction of the required inference window observed so far, and a period
convergence factor derived from the smoothed disagreement between the tracked
period and the current spectral estimate. Their product weights the predictive
term against a reactive fallback, so a model that has not yet settled degrades
gracefully to conventional guiding.

The INDIGO implementation reports the same two scalars to the operator as a
single \emph{learning} percentage, but \emph{averaged} rather than multiplied:
progress is then visible from the first guide cycle instead of reading zero
until both factors are simultaneously satisfied, which is what the product
used for blending would show. The blend weight that actually gates the
correction is still the product; the displayed percentage is a diagnostic
only, and the two agree once both scalars reach~1. Neither the original
model nor its PHD2 implementation exposes a comparable readout: there, the
guider blends in the prediction once enough data has accumulated, but gives
no numeric indication of how far that process has progressed.

\section{The multi-kernel extension}

\subsection{Model}

We replace the single periodic term in~\eqref{eq:base} with a sum over $N$
stages, indexed $s = 0 \dots N-1$, where $s = 0$ is the worm:
\begin{equation}
  k(d) \;=\;
  \sigma_0^2 e^{-d^2/2\ell_0^2}
  \;+\; \sum_{s=0}^{N-1} k_s(d)
  \;+\; \sigma_1^2 e^{-d^2/2\ell_1^2},
  \label{eq:mk}
\end{equation}
with each stage contributing
\begin{equation}
  k_s(d) \;=\; \sigma_s^2
  \exp\!\left(-\frac{2\sin^2(\pi d / P_s)}{\ell_s^2}\right)
  \cdot
  \underbrace{\exp\!\left(-\frac{d^2}{2\ell_{d,s}^2}\right)}_{\text{envelope, } s>0}.
  \label{eq:stage}
\end{equation}
\begin{table}[!htb]
  \centering
  \small
  \caption{Symbols used in \eqref{eq:base}--\eqref{eq:stage} and in the rest
  of the paper. Amplitudes are given in the natural parameterisation, in which
  the signal variance is the square of the tabulated quantity; length scales
  are likewise natural, a time, converted to the standard form of the periodic
  kernel by $\ell \mapsto 4\sin(\ell\pi/P)$. Defaults are those of the
  released implementation.}
  \label{tab:notation}
  \setlength{\tabcolsep}{5pt}
  \begin{tabular}{@{}l p{8.3cm} r@{}}
    \toprule
    Symbol & Meaning & Default \\
    \midrule
    $d = t - t'$      & lag between two points in gear time & --- \\
    $\beta$           & coefficients of the explicit linear trend & fitted \\
    $T$               & length of the inference window & --- \\
    \midrule
    $\sigma_0,\ \ell_0$ & amplitude and correlation time of the long SE term,
                            which carries slow aperiodic wander
                          & \SI{20}{\px}, \SI{700}{\second} \\
    $\sigma_1,\ \ell_1$ & the same for the short SE term, which absorbs
                            transient wander and is dropped from the projection
                          & \SI{10}{\px}, \SI{25}{\second} \\
    $\sigma_P,\ \ell_P,\ P$ & amplitude, length scale and period of the worm's
                            periodic term; $\ell_P$ sets how fine a feature is
                            resolved within one period, and hence how many
                            harmonics the term can carry
                          & \SI{20}{\px}, \SI{10}{\second}, tracked \\
    \midrule
    $N$               & number of periodic stages modelled & 2 \\
    $s$               & stage index; $s=0$ is the worm & --- \\
    $\sigma_s,\ \ell_s,\ P_s$ & the three quantities above, for stage $s$ & see $\kappa$, $\gamma_s$ \\
    $\kappa$          & ratio fixing $\ell_s = P_s/\kappa$ for $s>0$, so that a
                          stage's harmonic capacity does not depend on its period
                          & 13 \\
    $\ell_{d,s} = \lambda P_s$ & length scale of the decay envelope that makes
                          stage $s>0$ quasi-periodic & --- \\
    $\lambda$         & envelope length in units of the stage's own period & 20 \\
    \midrule
    $a_s$             & smoothed amplitude of stage $s$'s spectral line,
                          relative to the worm's & measured \\
    $a_\mathrm{off},\ a_\mathrm{full}$ & gate thresholds on $a_s$ & 0.15, 0.35 \\
    $\gamma_s$        & evidence gate for stage $s$, $0 \le \gamma_s \le 1$ & --- \\
    $\sigma_\mathrm{prior}$ & amplitude a fully gated-in stage is given & \SI{20}{\px} \\
    \midrule
    $\epsilon$        & relative error in a tracked period & --- \\
    $\Delta P$        & smallest separation two lines can be resolved at & \eqref{eq:rayleigh} \\
    \bottomrule
  \end{tabular}
\end{table}

The sum is the right composition. Two gear stages act in series on the same
shaft, so their contributions to the pointing error add; a \emph{product} of
periodic kernels would model amplitude modulation of one stage by another,
which is a different mechanism and not the one at issue here. Being a sum of
positive semi-definite kernels, \eqref{eq:mk} is itself positive semi-definite,
so no property of the inference is disturbed.

All stages remain in the projection kernel. A further gear stage is
predictable error, not wander, and omitting it from the forecast would defeat
the purpose of modelling it.

\subsection{Why a second kernel is necessary}

Let two stages have periods $P_0, P_1$ and consider the predictable signal
$g(t) = g_0(t) + g_1(t)$ with $g_s$ periodic at $P_s$. If $P_0/P_1 = p/q$ is
rational, $g$ is periodic at the common period $\mathrm{lcm}(P_0,P_1)$ (least
common multiple), and a single
periodic kernel at that period can represent it in principle. In practice this
is useless: the inference window would have to span the common period, which
for realistic ratios is many times longer than either stage, and the kernel
length scale would have to resolve $\max(p,q)$ harmonics.

If the ratio is irrational, no single periodic kernel represents $g$ at all.
The failure is not subtle---in our synthetic case C8 (Sec.~\ref{sec:synthetic})
the single-kernel model recovers almost none of the second stage, and the
residual is more than twice that of the multi-kernel model.

The decisive practical advantage of $N$ separate kernels over one kernel at the
common period is the data requirement. Warm-up scales with $\max_s P_s$, not
with the common period or the beat period $1/|1/P_0 - 1/P_1|$.

\subsection{Length scale tied to the period}
\label{sec:lengthscale}

The periodic kernel's length scale $\ell_s$ controls the finest feature
resolved within one period, and hence how many harmonics the stage can carry.
In the natural parameterisation the length scale is a time, converted to the
standard form by $\ell \mapsto 4\sin(\ell \pi / P)$. A length scale fixed in
absolute time therefore means something different for every period: at
\SI{33}{\second} it is a thirteenth of a \SI{432}{\second} primary but more
than a fifth of a \SI{155}{\second} stage, giving
$\exp(-0.5\sin^2(\pi d/P))$---a kernel so smooth it can represent little
beyond the fundamental. A second stage so parameterised is fitted as a bare sinusoid, and
most of the benefit of using a GP at all is lost.

We therefore set $\ell_s = P_s / \kappa$ with $\kappa = 13$ for $s>0$, which
keeps the harmonic capacity of a stage independent of how fast it happens to
be. The worm retains its absolute length scale, for backward compatibility with
tuning established for the single-kernel model.

\subsection{Quasi-periodic envelope}
\label{sec:envelope}

A strictly periodic kernel is coherent over the whole inference window: an
observation one hundred cycles ago constrains the prediction as strongly as one
a single cycle ago. That is precisely what makes it powerful, and also what
makes it fragile to period error. If the tracked period is wrong by a relative
amount $\epsilon$, the accumulated phase error across a window spanning $n$
cycles is $n\epsilon$ cycles. Since $n = T/P_s$ for a window of length $T$, a
short stage accumulates error faster than a long one in exact proportion to its
frequency: a \SI{10800}{\second} window holds 25 cycles of a
\SI{432}{\second} primary but 70 of a \SI{155}{\second} stage, so the same
$\epsilon$ produces nearly three times the phase error.

\begin{figure}[!htb]
  \centering
  \includegraphics[width=0.62\textwidth]{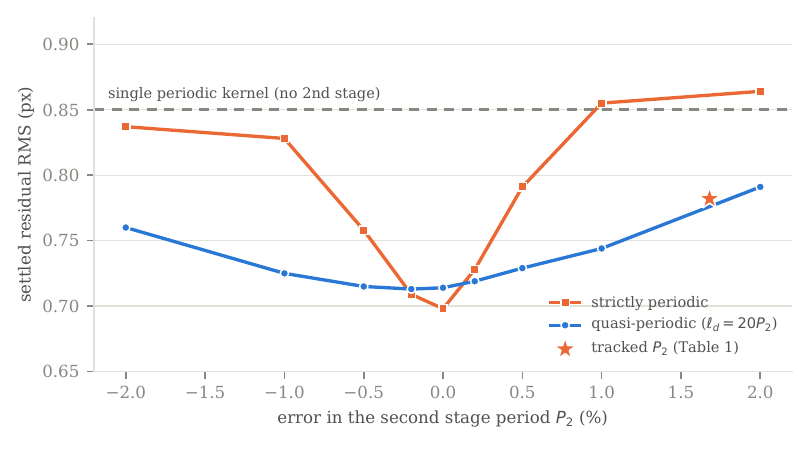}
  \caption{Sensitivity of the second periodic kernel to error in its period
  (synthetic case C5). Both periods are pinned here, since $P_2$ has to be held
  in order to be swept; the values are therefore lower than the corresponding
  entry of Table~\ref{tab:synth}, where both periods are tracked online. A
  strictly periodic kernel is only useful within about
  $\pm\SI{0.2}{\percent}$; the quasi-periodic form with $\ell_d = 20P_2$
  degrades far more gently. The dashed line is the single-kernel model. The
  star marks where the online estimator actually lands when $P_2$ is tracked
  rather than pinned: it settles at \SI{157.6}{\second}, an error of
  \SI{1.7}{\percent}, and the residual it achieves is close to what the curve
  predicts at that error.}
  \label{fig:sens}
\end{figure}

This is not a hypothetical. Figure~\ref{fig:sens} (orange) shows the closed-loop
residual for synthetic case C5 as the pinned second period is varied. At the
exact period the second kernel reduces the residual from \SI{0.850}{\px} to
\SI{0.698}{\px}; at $\pm\SI{1}{\percent}$ all of that gain is gone, and by
$+\SI{2}{\percent}$ the kernel is worse than useless. An FFT estimate good to
\SI{0.5}{\percent}---entirely typical---already costs \SI{0.09}{\px}.

The remedy is to make the extra stages quasi-periodic by multiplying the
periodic core with a squared-exponential envelope, as in~\eqref{eq:stage}, with
$\ell_{d,s} = \lambda P_s$. The envelope limits how many cycles back the
prediction draws on, so phase error ceases to accumulate; the cost is that less
data is averaged. Figure~\ref{fig:sens} (blue) shows the result for
$\lambda = 20$: the residual degrades far more gently---at
$\pm\SI{1}{\percent}$ it is \SI{0.73}{\px} against \SI{0.84}{\px} for the
strictly periodic form---at the price of \SI{0.016}{\px} at the exact period.

Figure~\ref{fig:sens} also quantifies what the online period estimator costs.
Pinning both periods at their true values gives \SI{0.714}{\px} on this case,
while tracking them---the configuration a user obtains---gives
\SI{0.782}{\px}. The tracker settles at \SI{157.6}{\second} against a true
\SI{155}{\second}, an error of \SI{1.7}{\percent}, and reading that error off
the curve accounts for almost the whole difference.

The last two columns of Table~\ref{tab:synth} give this comparison for every
case, and the cost is strikingly uneven. Where the second line is clean and
well separated it is negligible---\SI{0.001}{\px} on both C3 and
C8---but where it sits among the primary's harmonics it dominates:
\SI{0.068}{\px} on C5, \SI{0.150}{\px} on C6 and \SI{0.077}{\px} on C7. On
precisely those cases the estimator's residual bias, not the covariance, is
what limits the second stage, and a better period estimate would be worth more
than any further change to the kernel.
Shorter envelopes cost accuracy without buying further robustness, so
$\lambda = 20$ is the knee.

\subsection{Discovering further stages}
\label{sec:discovery}

Stage $s>0$ is found from the same spectrum used for the worm. Let
$\mathcal{K} = \{P_0,\dots,P_{s-1}\}$ be the periods already resolved. The
procedure is:

\begin{enumerate}
\item Compute the power spectrum of the de-trended, Hamming-windowed,
      regularised residual. Note that this returns $|X|^2$; the strength
      measure below is an \emph{amplitude} ratio and must take the square root.
\item Notch out every $P \in \mathcal{K}$, its harmonics $kP$ for
      $k \le 5$, and its first sub-harmonics $P/k$ for $k \in \{2,3\}$, each
      over a relative width of \SI{8}{\percent} in frequency.
\item Take the strongest surviving line, confined either to a band around a
      user-supplied seed, or---unseeded---to
      $[4\Delta_{\text{grid}},\, 1.2 P_0]$. The upper bound matters: anything
      slower than the worm is drift, already carried by the long SE kernel and
      the linear trend, and without the bound the estimator intermittently
      reports a \SIrange{300}{350}{\second} ``stage'' that is only the residual
      of an imperfect de-trend.
\item Reject the candidate if it is commensurate with any $P \in \mathcal{K}$,
      that is if $|P/(kP_s) - 1| < 0.06$ or $|P_s/(kP) - 1| < 0.06$ for any
      $k \le 5$.
\end{enumerate}

The commensurability guard in step 4 is essential. A harmonic of an
already-modelled line is already represented by that stage's length scale;
admitting it as a separate stage makes the two terms degenerate---they explain
the same signal, the split of variance between them is arbitrary, the Gram
matrix conditioning degrades, and the two period trackers oscillate against one
another.

\subsection{Evidence gating}

A stage that is enabled is not thereby used. Its signal variance is scaled by a
gate driven by the smoothed amplitude $a_s$ of its spectral line relative to the
worm's:
\begin{equation}
  \sigma_s^2 \;=\; \gamma_s\,\sigma_\mathrm{prior}^2, \qquad
  \gamma_s = \mathrm{clip}\!\left(\frac{a_s - a_\mathrm{off}}{a_\mathrm{full} - a_\mathrm{off}},\,0,\,1\right),
\end{equation}
with $a_\mathrm{off} = 0.15$ and $a_\mathrm{full} = 0.35$. Below the lower
threshold the stage is switched out entirely, so a mount with a single periodic
component behaves exactly as under the single-kernel model. This is what makes
the extension safe to enable by default on unknown hardware, and it is the
property we regard as most important for a control loop that runs unattended.

Finally, the period convergence factor of Sec.~\ref{sec:background} is extended
to take the worst over all stages that are both tracking their period and
carrying real weight: the prediction is only as trustworthy as the least
settled period it rests on.

\section{Implementation}

The model is implemented in dependency-free, multiplatform code in the C
programming language, as part of the INDIGO distribution~\cite{indigo}. The
dense linear algebra (LDLT factorisation and
solve) and a radix-2 FFT are implemented directly, so no external library is
required. The hyperparameter vector grows from 7 to $7 + 3N$ entries; the
kernel, the estimator, the gate and the convergence factor all iterate over
stages, so raising $N$ is a one-line change, although only $N=2$ is validated
here. Stage 0 is the worm: it is always modelled and never gated, so
$\gamma_0 = 1$ by construction.

Two files carry the work, both given relative to the root of the INDIGO
source tree\footnote{\url{https://github.com/indigo-astronomy/indigo}}:

\begin{itemize}\setlength{\itemsep}{2pt}
\item[] \texttt{indigo\_libs/indigo\_gp\_guider.c}\\
  the guiding algorithm itself---the covariance of
  \eqref{eq:mk}--\eqref{eq:stage}, the spectral search of
  Sec.~\ref{sec:discovery}, the evidence gate and the period trackers, with
  its public interface in the matching header. It has
  no dependency on the rest of INDIGO beyond a logging call, and can be
  compiled and driven standalone, which is how every measurement in this paper
  was produced.
\item[] \texttt{indigo\_drivers/agent\_guider/indigo\_agent\_guider.c}\\
  the wiring to the mount and to the user: the multi-kernel model appears there
  as a selectable right-ascension correction mode alongside the single-kernel
  one, with the second stage's period and tracking exposed as ordinary agent
  properties.
\end{itemize}

Measured cost is \SIrange{0.53}{0.69}{\milli\second} per guide frame on an
Apple M-series core with 100 inducing points, dominated by the second FFT;
against a multi-second guide cycle this is negligible.

\section{Synthetic evaluation}
\label{sec:synthetic}

\subsection{Method}

\begin{figure}[!htb]
  \centering
  \includegraphics[width=0.78\textwidth]{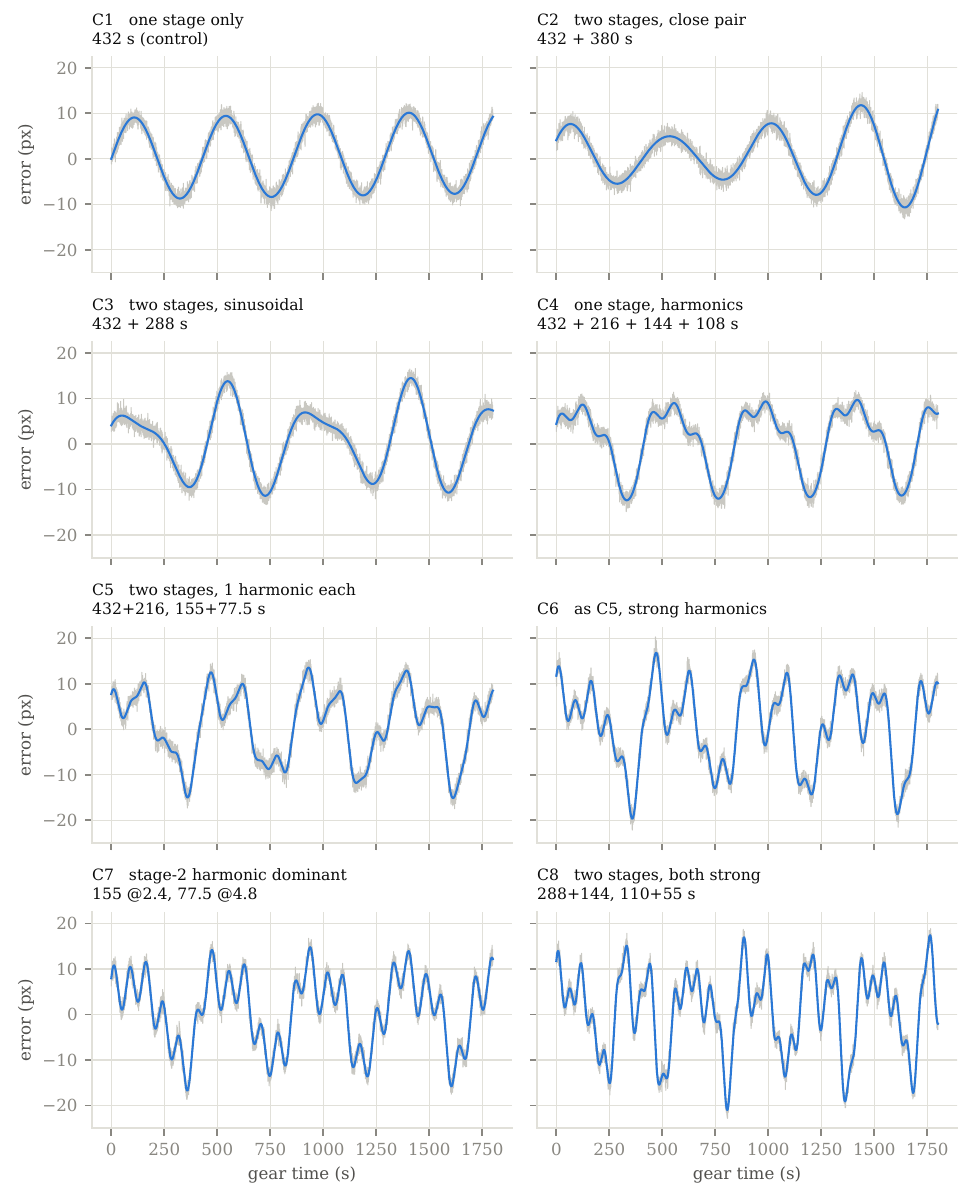}
  \caption{The eight synthetic gear-error waveforms, shown over
  \SI{1800}{\second}, a little over four cycles of the \SI{432}{\second}
  primary. In each panel the upper trace (blue) is the true error and the lower
  trace (grey) is the same error as the controller measures it, with the
  \SI{1.0}{\px} of Gaussian measurement noise used throughout
  Sec.~\ref{sec:synthetic}; the two are drawn with separate zeros so both stay
  legible. C1 and C4 are single-stage---C4's extra lines are exact harmonics of
  the primary---while the rest add a genuine second stage that is not a
  harmonic of the first. C1--C4 are ordered by how far the second line sits
  from the primary.}
  \label{fig:curves}
\end{figure}

We drive the controller in a closed loop against a synthetic gear error
composed of sinusoids plus a linear drift, with additive Gaussian measurement
noise of \SI{1.0}{\px}---and, for comparison, without noise at all---and a
guide cycle of \SI{2}{\second} over
\SI{10800}{\second} (\SI{3}{\hour}). Periods and amplitudes are chosen to
match commercial hardware: a primary stage of \SI{432}{\second} at
\SI{9}{\px} amplitude, which at a typical image scale corresponds to the
\SIrange{20}{25}{\arcsec} peak-to-peak seen on the strain-wave mounts of
Sec.~\ref{sec:real}. Both matter: scaling the period without scaling the
amplitude reduces the slew rate the controller must follow, and the comparison
collapses into the noise floor. At each frame the controller observes the gear error plus
everything it has already applied, and the residual is scored over the final
two thirds of the run, by which point the model has learned. Eight cases, shown
in Fig.~\ref{fig:curves}, span the
regimes of interest, ordered by how far the second line sits from the primary:
a single sinusoidal stage as control (C1); a close pair at \num{432} and
\SI{380}{\second} (C2); two sinusoidal stages at \num{432} and
\SI{288}{\second} (C3), a 3:2 ratio in which neither is a harmonic of the
other, so both kernels are genuinely required; a single stage whose extra lines
are exact harmonics (C4); two well-separated stages each with a harmonic
(C5, C6); a second stage whose harmonic dominates its own fundamental (C7);
and two strongly non-sinusoidal stages at \num{288} and \SI{110}{\second},
both carrying a strong second harmonic (C8).

\subsection{Results}

\begin{figure}[!htb]
  \centering
  \includegraphics[width=\textwidth]{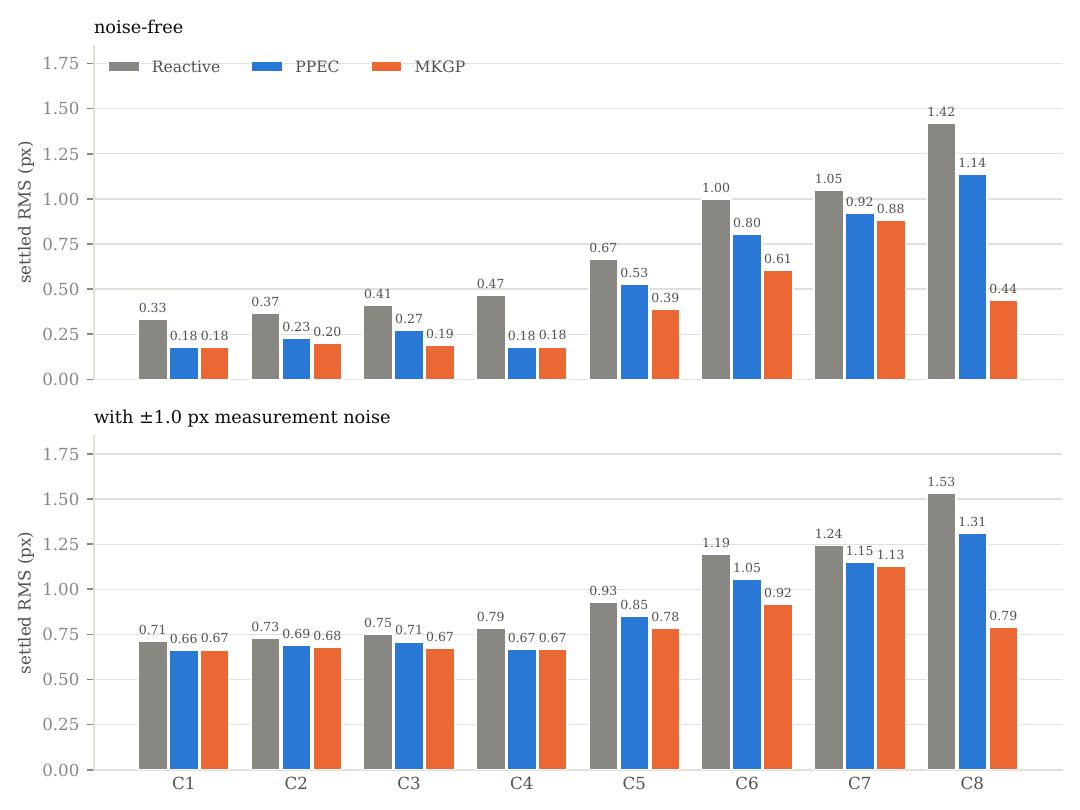}
  \caption{Settled residual RMS on the synthetic cases. The multi-kernel model
  is unchanged from the single-kernel model on the single-stage controls
  (C1, C4) and reduces the residual by \SIrange{1}{40}{\percent} wherever a
  genuine second stage is present, the gain growing with how well separated the
  two stages are.}
  \label{fig:results}
\end{figure}

Table~\ref{tab:synth} and Fig.~\ref{fig:results} give the settled residual RMS
for both noise levels. Without noise the multi-kernel model reduces the
residual against the single-kernel one by \SI{12}{\percent} (C2),
\SI{31}{\percent} (C3), \SI{26}{\percent} (C5), \SI{25}{\percent} (C6) and
\SI{62}{\percent} (C8). With \SI{1.0}{\px} of noise the same figures become
\SI{1}{\percent}, \SI{5}{\percent}, \SI{8}{\percent}, \SI{13}{\percent} and
\SI{40}{\percent}.

Equally important, the two single-stage controls are unchanged at both noise
levels: C4 differs by \SI{0.001}{\px} and C1 not at all without noise, and by
\SI{0.000}{\px} and \SI{0.001}{\px} with it. In both the gate correctly
switched the second stage out---in C4 because every candidate line
(\SI{216}{}, \SI{144}{}, \SI{108}{\second}) is a harmonic of the primary and
was refused by the commensurability guard, and in C1 because no second line
exists.

\subsection{How measurement noise compresses the comparison}
\label{sec:noise}

The two halves of Table~\ref{tab:synth} differ in a way that is worth stating
explicitly, because it governs how the synthetic results should be read against
a real mount. Noise does not change the ranking of the three controllers in any
case, nor does it change which cases the second kernel helps on. What it does
is add a nearly constant term in quadrature. Taking the difference of squares
between the two halves gives \SIrange{0.64}{0.70}{\px} for every case and
every controller---a spread of \SI{9}{\percent} across conditions whose
noise-free residuals span a factor of eight. The loop passes a little under
two-thirds of the \SI{1.0}{\px} input noise through to the residual, and it
does so almost independently of what the gear error looks like or which model
is correcting it.

The consequence is that noise compresses relative differences without removing
them. Where the noise-free residual is small the added term dominates: C3 falls
from a \SI{31}{\percent} advantage to \SI{5}{\percent}, and C1 and C4 sit at
\SIrange{0.66}{0.67}{\px} for all three controllers because \SI{0.64}{\px} of
that is noise the controller cannot touch. Where the gear error is large enough
to stand above the added term the advantage survives largely intact: C8 keeps
\SI{40}{\percent} of its \SI{62}{\percent}. A mount therefore benefits from
the second kernel in proportion to how far its periodic error rises above the
guiding noise floor, which is a property of the mount and the guide star
together rather than of the algorithm.

It also means the noise level chosen here is a pessimistic one. At
\SI{1.0}{\px} against a \SI{18}{\px} peak-to-peak primary the synthetic cases
sit at a signal-to-noise ratio of about \num{18}, against about \num{34} for
the raw log of the mount of Sec.~\ref{sec:real} before de-noising. The noisy
half of Table~\ref{tab:synth} should be read as a lower bound.

C7 deserves comment, because it overturns an assumption we started with. Here
the second stage's harmonic (\SI{77.5}{\second}, amplitude \SI{4.8}{\px}) is
louder than its own fundamental (\SI{155}{\second}, \SI{2.4}{\px}). Seeded at
the fundamental, as in Table~\ref{tab:synth}, the gate barely opens and the
model gains little. Left to search unseeded, it locks onto the
\SI{77.5}{\second} line and reaches \SI{0.604}{\px}, a \SI{37}{\percent}
improvement---substantially better than seeding the fundamental.

This is the opposite of what we expected. The intuition that a kernel belongs
on the fundamental, because it then represents the harmonics too, is sound in
principle but loses to the simpler fact that a kernel should be placed where
the energy is. We had implemented the corresponding remedy---a rule
reconstructing the fundamental from a harmonic, analogous to the one the
primary estimator already uses---before measuring this, and it proved
\emph{net negative} for a second reason as well: it cost
\SIrange{0.045}{0.096}{\px} across the other two-stage cases through false
positives on the leakage skirt of the notched primary line, to gain little
anywhere. Three mitigations (an amplitude floor, a local-maximum test, and a
persistence vote) failed to separate the good firings from the bad, because the
false positives are genuine local maxima and arrive in bursts. The rule is
therefore not included, and the unseeded search is left to follow the strongest
admissible line.

C8 is the clearest demonstration in the suite, and the case closest to what a
two-stage strain-wave mount actually produces: two independent stages, both
strong and both markedly non-sinusoidal. A single periodic kernel can follow
one harmonic family but not two, and the residual falls from \SI{1.312}{\px}
to \SI{0.790}{\px}, a reduction of \SI{40}{\percent}. The unseeded search
performs almost as well (\SI{0.839}{\px} against \SI{1.318}{\px}, a
\SI{36}{\percent} reduction), locking onto \SI{110.0}{\second} at full weight
without being told where to look.

A variant of this case is worth recording as a limitation. Moving the second
stage from \num{110} to \SI{150}{\second}, so that the ratio $288/150 = 1.92$
falls within \SI{4}{\percent} of 2:1, suppresses it entirely: the
commensurability guard of Sec.~\ref{sec:discovery} evaluates
$|P_0/(2P_s) - 1| = 0.04$ and refuses the candidate, and independently the notch over the primary's own second harmonic
at \SI{144}{\second} spans \SIrange{132}{157}{\second} and erases the line
before any search runs. The gate never opens and the model reduces to the
single-kernel one. That the stage is real is easily shown---a narrowband
canceller outside the GP, using neither notch nor guard, recovers it---so the
limitation lies in the discovery procedure rather than the model. A stage lying
near a small-integer ratio of one already present is, for this estimator,
indistinguishable from that stage's own harmonic. Narrowing the notch and the
guard tolerance would resolve it at the cost of reintroducing the degeneracy
they exist to prevent, which is the commoner failure.

\begin{table}[t]
  \centering
  \small
  \caption{Settled residual RMS (px) on the synthetic cases, run without
  measurement noise and again with \SI{1.0}{\px} of it. Open loop is the
  uncorrected gear error; reactive is a hysteresis controller. Every period is
  estimated online, as a user would obtain it.}
  \label{tab:synth}
  \setlength{\tabcolsep}{4pt}
  \begin{tabular}{llrrrrrrr}
    \toprule
    & & & \multicolumn{3}{c}{noise-free} & \multicolumn{3}{c}{\SI{1.0}{\px} noise} \\
    \cmidrule(lr){4-6}\cmidrule(l){7-9}
    Case & Description & Open loop & React. & PPEC & MKGP & React. & PPEC & MKGP \\
    \midrule
    C1 & one stage only (control)    & 8.634 & 0.332 & \textbf{0.176} & \textbf{0.176} & 0.712 & \textbf{0.664} & 0.665 \\
    C2 & two stages, close pair      & 9.028 & 0.367 & 0.226 & \textbf{0.198} & 0.730 & 0.689 & \textbf{0.681} \\
    C3 & two stages, sinusoidal      & 9.166 & 0.410 & 0.271 & \textbf{0.187} & 0.753 & 0.709 & \textbf{0.673} \\
    C4 & one stage + harmonics       & 9.019 & 0.468 & \textbf{0.180} & 0.181 & 0.786 & \textbf{0.670} & \textbf{0.670} \\
    C5 & two stages, 1 harmonic each & 9.478 & 0.666 & 0.525 & \textbf{0.390} & 0.927 & 0.850 & \textbf{0.782} \\
    C6 & as C5, strong harmonics     & 10.436 & 0.998 & 0.804 & \textbf{0.606} & 1.194 & 1.055 & \textbf{0.919} \\
    C7 & stage-2 harmonic dominant   & 9.661 & 1.050 & 0.920 & \textbf{0.880} & 1.243 & 1.150 & \textbf{1.126} \\
    C8 & two stages, both strong     & 10.604 & 1.421 & 1.137 & \textbf{0.438} & 1.534 & 1.312 & \textbf{0.790} \\
    \bottomrule
  \end{tabular}
\end{table}

\section{Recorded data}
\label{sec:real}

\subsection{Method}

A recorded session is replayed in closed loop: each row of the log supplies one
guide frame, its elapsed time sets the guide cycle, and its reconstructed
open-loop periodic error is the gear error. No synthetic noise is added,
because the recorded error already carries the measurement noise of the
session. The residual is scored over the second half of the session, by which
point both GP models have learned. We present one mount here, the MLAstro
SAL-33 of Sec.~\ref{sec:sal33}; it is a single example and is not offered as a
survey of hardware.

We report two operating points: the shipped default gain of each controller,
which is what a user obtains without tuning, and a raised setting that leaves
the reactive baseline at its default while giving both GP models
\SI{70}{\percent} reactive and \SI{60}{\percent} predictive gain. Comparing
at a single arbitrary gain would be misleading, because the shipped defaults do
not sit at the same point on each controller's gain curve.

\subsection{MLAstro SAL-33, de-noised}
\label{sec:sal33}

\begin{figure}[!htb]
  \centering
  \includegraphics[width=0.95\textwidth]{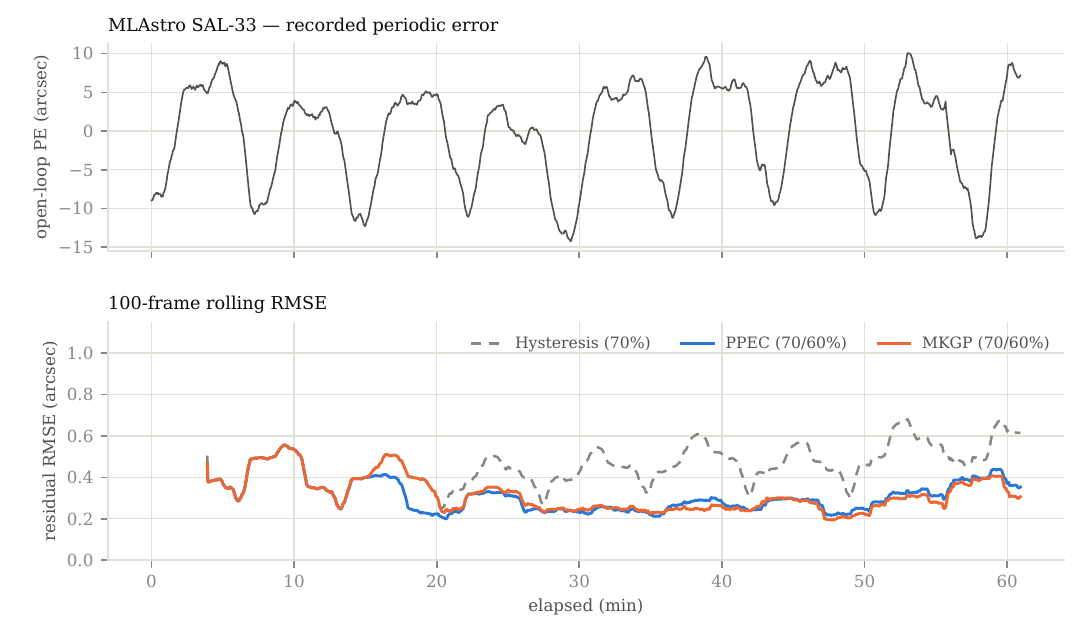}
  \caption{A recorded, de-noised session on an MLAstro SAL-33. Top: the
  reconstructed open-loop periodic error. Bottom: 100-frame rolling residual
  RMSE for the three controllers, hysteresis at \SI{70}{\percent}
  aggressiveness and both GP models at \SI{70}{\percent} reactive /
  \SI{60}{\percent} predictive gain. The first
  rolling measurement of each series, which covers the interval in which the GP
  models are still acquiring, is omitted. Figures quoted in the text are scored
  over the second half of the session.}
  \label{fig:real}
\end{figure}

Figure~\ref{fig:real} shows a \SI{1.06}{\hour} session at \SI{2.4}{\second}
cadence recorded on an MLAstro SAL-33, a mount whose error is dominated by a
\SI{428}{\second} line with a strong second harmonic at \SI{217}{\second},
\SI{24}{\arcsec} peak to peak. The version used here has been de-noised,
leaving a residual measurement noise of \SI{0.07}{\arcsec} against
\SI{0.71}{\arcsec} in the raw log; this isolates tracking performance from the
measurement floor, which otherwise dominates. The first \SI{169}{\second} of
the log, an acquisition transient in which the error falls monotonically
through \SI{24}{\arcsec} at roughly twice the slope of any subsequent cycle,
have been removed; the record begins at the first trough, leaving
\SI{1.02}{\hour}.

With hysteresis at its default aggressiveness of \SI{70}{\percent} and both GP
models at \SI{70}{\percent} reactive and \SI{60}{\percent} predictive gain, the
residual RMSE over the second half of the session is \SI{0.513}{\arcsec} for
hysteresis, \SI{0.303}{\arcsec} for the single-kernel model and
\SI{0.284}{\arcsec} for the multi-kernel model: an improvement of
\SI{40.9}{\percent} and \SI{44.6}{\percent} respectively over the reactive
baseline, and \SI{6.3}{\percent} of the multi-kernel model over the
single-kernel one. Table~\ref{tab:sal33} collects these figures together with
the shipped defaults; note that the advantage of the multi-kernel model over
the single-kernel one is essentially independent of the reactive gain, while
its advantage over the reactive baseline is not.

\begin{table}[t]
  \centering
  \small
  \caption{Settled residual RMSE (\si{\arcsec}) on the trimmed, de-noised MLAstro
  SAL-33 session, at two operating points. Open loop is
  \SI{6.645}{\arcsec}. ``Defaults'' is each controller's shipped gain
  (hysteresis \SI{70}{\percent} aggressiveness; GP models \SI{60}{\percent}
  reactive, \SI{50}{\percent} predictive). ``Raised'' leaves hysteresis at its
  default and puts both GP models at \SI{70}{\percent} reactive,
  \SI{60}{\percent} predictive.}
  \label{tab:sal33}
  \begin{tabular}{lcc}
    \toprule
    Controller & Defaults & Raised \\
    \midrule
    Hysteresis           & 0.513 & 0.513 \\
    PPEC (single kernel) & 0.356 & 0.303 \\
    MKGP (multi kernel)  & \textbf{0.334} & \textbf{0.284} \\
    \midrule
    MKGP vs.\ hysteresis & $-34.9\%$ & $-44.6\%$ \\
    MKGP vs.\ PPEC       & $-6.2\%$  & $-6.3\%$  \\
    \bottomrule
  \end{tabular}
\end{table}

\subsection{Why the gain is small}

\begin{figure}[!htb]
  \centering
  \includegraphics[width=\textwidth]{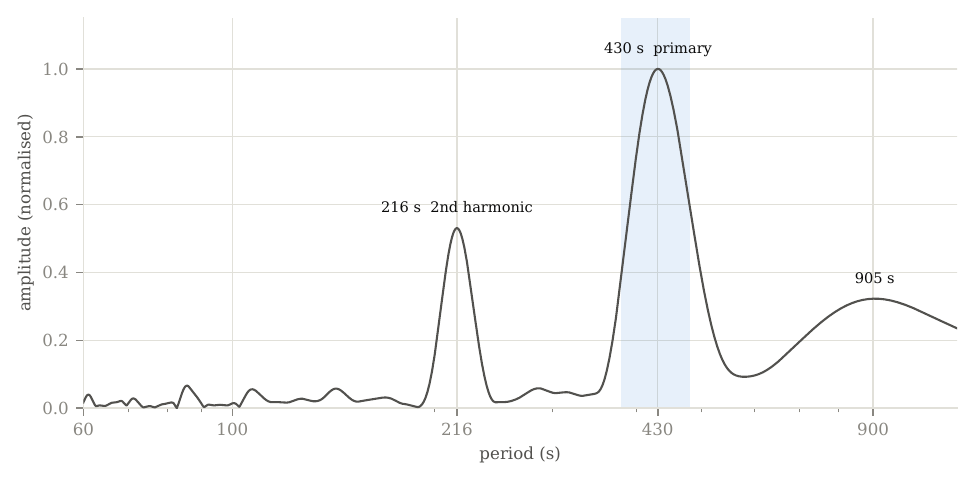}
  \caption{Amplitude spectrum of the MLAstro SAL-33 curve of
  Fig.~\ref{fig:real}, Hann-windowed and zero-padded. Three features are
  present: the \SI{430}{\second} primary, its own second harmonic, and a
  low-frequency residue of the de-trend. There is no second gear stage. The
  shaded band is the resolution limit \eqref{eq:rayleigh} about the primary:
  no second line inside it could be separated from the primary by a record of
  this length.}
  \label{fig:spectrum}
\end{figure}

The improvement of the multi-kernel model over the single-kernel one on this
session---\SI{6.3}{\percent}---is far smaller than on the synthetic two-stage
cases, and the spectrum in Fig.~\ref{fig:spectrum} explains why. Only three
lines rise above the noise: the primary at \SI{430}{\second}, its own second
harmonic at \SI{216}{\second}, and a broad low-frequency feature near
\SI{900}{\second} that is the residue of the de-trend. There is no second
independent gear stage. A single periodic kernel, whose length scale already
lets it carry harmonics, covers this spectrum in full.

No second gear stage is present, so there is nothing of that kind for a second
kernel to model. Nor could a second stage close to the primary have been
identified from a record of this length: the frequency resolution of a record
of length $T$ is $\Delta f = 1/T$, or in period terms
\begin{equation}
  \Delta P \;=\; \frac{P^2}{T},
  \label{eq:rayleigh}
\end{equation}
which for $P = \SI{430}{\second}$ and $T = \SI{3655}{\second}$ gives
\SI{51}{\second}: any second line within that of the primary is, as far as
this record is concerned, the same line.

Two separate things then account for the result, and neither is a second gear
stage. The first is that most of what the single-kernel model leaves behind is
not periodic at all. Partitioning the power by period band, the single-kernel
model removes the primary almost completely---the
\SIrange{300}{600}{\second} band containing it falls from \SI{69}{\percent} of
the total to \SI{11}{\percent}---but \SI{84}{\percent} of what remains now
lies below \SI{300}{\second}, where the open-loop error had essentially
nothing (\SI{0.1}{\percent} below \SI{60}{\second}). That content is generated by the
control loop itself: the one-frame lag, the reactive term chasing its own
corrections, and the discrete cadence of \SI{2.4}{\second}. No periodic kernel
addresses any of it, which is why the single-kernel model plateaus near
\SI{0.3}{\arcsec} with the periodic part of the problem already solved. The
effect is the same one Sec.~\ref{sec:noise} isolates on the synthetic cases: a
contribution the controller cannot remove sets a floor, and differences between
models are compressed against it.

The second is that the primary is not stationary, and this is where the
\SI{6.3}{\percent} comes from. Fitting a \SI{430}{\second} sinusoid to each
third of the session gives amplitudes of \SIlist{7.72;7.03;8.79}{\arcsec} and
phases of \SIlist{-157.5;-161.3;-164.5}{\degree}: the amplitude varies by about
\SI{11}{\percent} and the phase creeps by \SI{7}{\degree} over the hour. A
strictly periodic kernel is coherent across the whole inference window, so it
fits one average cycle and cannot follow a primary whose amplitude is growing.
The extra kernel sits close to the primary and carries a decay envelope, and
two close periods beating over about one session length are indistinguishable
from a single period with a slowly varying amplitude---which is precisely what
the mount is doing. The second kernel is therefore modelling amplitude
modulation of the primary rather than a second stage. That is a real effect,
which is why the improvement is real and repeatable across gains
(Table~\ref{tab:sal33}); it is a second-order effect, which is why it is only
\SI{6.3}{\percent}.

This sets a practical requirement that is easy to state and easy to overlook:
to identify a second stage at period $P_s$ the session must be long enough that
$T > P^2/|P - P_s|$, which is \eqref{eq:rayleigh} rearranged, and which for two
stages a few per cent apart means hours, not an hour. Where the synthetic cases separate cleanly---C8's stages at
\num{288} and \SI{110}{\second} differ by far more than
$\Delta P$---the model recovers the second stage and returns
\SI{40}{\percent}.

\section{Conclusions}

Extending GP-based predictive periodic error correction from one periodic
kernel to a sum over several is straightforward as a modelling step and
delivers roughly a halving of the residual wherever a genuine second gear stage
exists. Making it usable required three measures that are not obvious from the
model alone: the length scale of each stage must scale with its own period
(Sec.~\ref{sec:lengthscale}), or the stage degenerates to a sinusoid; the extra
stages must be quasi-periodic rather than strictly periodic
(Sec.~\ref{sec:envelope}), or they demand a period accuracy no online
estimator delivers; and each stage must be gated on spectral evidence and
refused when commensurate with a stage already present, or the model
double-counts a harmonic and degrades.

The resulting model reduces exactly to the published single-kernel model when
no second stage is present, which we regard as the precondition for enabling it
in an unattended control loop. The principal open item is a resolution guard on
the discovery procedure, rejecting a candidate unless the observed span covers
several beat periods against the stages already modelled; without one, a short
session can admit a second stage that is only the main lobe of the first.

\section*{Reproducibility}

The implementation is \texttt{indigo\_libs/\allowbreak indigo\_gp\_guider.c}
of the INDIGO distribution\footnote{\url{https://github.com/indigo-astronomy/indigo}}, as
released: every measurement in this paper was
produced against it with the default constants ($\kappa = 13$,
$\lambda = 20$, gate thresholds $0.15/0.35$, commensurability tolerance
$0.06$ for $k \le 5$), with no local modification. The recorded guiding logs
are available from the author on request.

\end{document}